# Identification of socioeconomic factors influencing global food price security using machine learning

Shan Shan[a]

[a]Sociology Department, Zhejiang University, Hangzhou 310058, China

Correspondence: shshan@zju.edu.cn

[1]Abbreviations

**Abstract:** Global concern over food prices and security has been exacerbated by the impacts of armed conflicts such as the Russia–Ukraine War, pandemic diseases, and climate change. Traditionally, analyzing global food prices and their associations with socioeconomic factors has relied on static linear regression models. However, the complexity of socioeconomic factors and their implications extend beyond simple linear relationships. By incorporating determinants, critical characteristics identification, and comparative model analysis, this study aimed to identify the critical socioeconomic characteristics and multidimensional relationships associated with the underlying factors of food prices and security. Machine learning tools were used to uncover the socioeconomic factors influencing global food prices from 2000 to 2022. A total of 105 key variables from the World Development Indicators and the Food and Agriculture Organization of the United Nations were selected.

---

[1] FAO  Food and Agriculture Organization
GNI  Gross national income
ODA  Official development assistance
SVR  Support vector regression
WDI  World Development Indicators

Machine learning identified four key dimensions of food price security: economic and population metrics, military spending, health spending, and environmental factors. The top 30 determinants were selected for feature extraction using data mining. The efficiency of the support vector regression model allowed for precise prediction-making and correlation analysis.

## 1. Introduction

Amidst the rapidly changing global landscape, rising food prices and security issues have emerged as prominent challenges. These concerns arise from a complex interplay of factors such as armed conflicts, large-scale health crises, and relentless climate change and require in-depth analysis and inventive solutions. The urgency has been made even clearer by recent global events, such as the Russia–Ukraine conflict and COVID-19 pandemic. The Food and Agriculture Organization (FAO) of the United Nations Food Price Index (FFPI, https://www.fao.org/prices/en/) provides insights into the changing prices of food on the global market. It is a useful tool for monitoring and analyzing trends and fluctuations in food prices that can have significant impacts on food security, trade, and agricultural policy.

Traditional research methods often use linear regression models to evaluate these issues. While these models have contributed significantly to our understanding of food prices and security, they tend to simplify the multiple interactions between phenomena and their many socioeconomic determinants. The impact of recent events has increased the vulnerability of global food systems. Disrupted supply chains, pressure on agricultural productivity, and rising inflation have led to dramatic fluctuations in food prices, threatening global food security. These challenges highlight the urgent need for advanced data-driven methods to examine the complex determinants of food prices and security. The complexity of different socioeconomic systems often contradicts simple linear assumptions and requires more advanced analytical approaches (Bar-Yam, 2004). Machine learning methods are particularly useful for elucidating the complex interplay of systems, especially for social phenomena (Grimmer et al., 2021). When coupled with feature extraction techniques such as principal component analysis, machine learning can improve the efficiency of data analysis by identifying the most influential variables (Awan et al., 2019). This not only reduces the number of variables but also helps to clarify the dynamics of complex systems (Hall and Smith, 1998).

The present study examined the socioeconomic aspects of global food prices and security. By harnessing the power of machine learning, this study aimed to overcome the limitations of traditional linear models and capture the complex and often non-linear relationships between a variety of variables that influence global food prices and security.

Specifically, this study aimed to determine to what extent machine learning algorithms capture the complex interactions among socioeconomic determinants of global food prices and security compared to traditional linear regression models, the specific socioeconomic factors that have had the greatest impact on global food prices and security in recent decades, and how to translate insights gained from machine learning models into actionable policy interventions to stabilize food prices and improve global food security, particularly considering local adaptations.

The overall goal was to illuminate the complexity of these issues and promote a comprehensive understanding of data-driven policymaking in this area. This study provides policymakers with robust strategies to address the impact of food price fluctuations and security issues on a global scale. Moreover, the methodological framework lays the foundation for effective mitigation strategies and emphasizes the need for local adaptation and future research in this area.

## 2. Methods

*2.1 Data*

The target variable was the food prices and security indicator derived from the FFPI (2014–2016 = 100)[ The FAO Food Price Index (FFPI) https://www.fao.org/worldfoodsituation/foodpricesindex/en/ ]. A total of 104 features were selected from the World Bank [World Development Indicators (WDI) https://datatopics.worldbank.org/world-development-indicators/] database as described in Supplementary Table S1. Political and military indicators included armed forces personnel (total number and percentage of the total labor force) and military expenditure (percentage of GDP). Economic indicators included foreign direct investment (net outflows and inflows, percentage of GDP); urban population (and its annual growth); agriculture, forestry, and fisheries (value added as a percentage of GDP); services and trade (value added as a percentage of GDP); GDP per capita (constant 2015 US dollar and its annual growth); manufacturing value added; manufacturing exports (percentage of goods exports); exports of goods and services, goods and services expense (percentage of total expenditure); and urban population (percentage of total population). Health indicators included life expectancy at birth, lifetime risk of maternal death (percentage and 1 in "x" rate), health spending measures such as current health spending (percentage of GDP and per capita in US and international dollars adjusted to purchasing power parity), and domestic general government health expenditure (percentage of GDP). Taken together, these indicators provide a comprehensive overview of global food prices, security, and related societal factors.

Food prices between 2000 and 2022 showed a generally increasing trend with significant fluctuations, particularly between 2006 and 2011. The period after 2011 showed more stability, albeit at higher price levels than in the early 2000s. This pattern highlights the dynamics of food prices, which are influenced by a range of socioeconomic and environmental factors (Fig. 1).

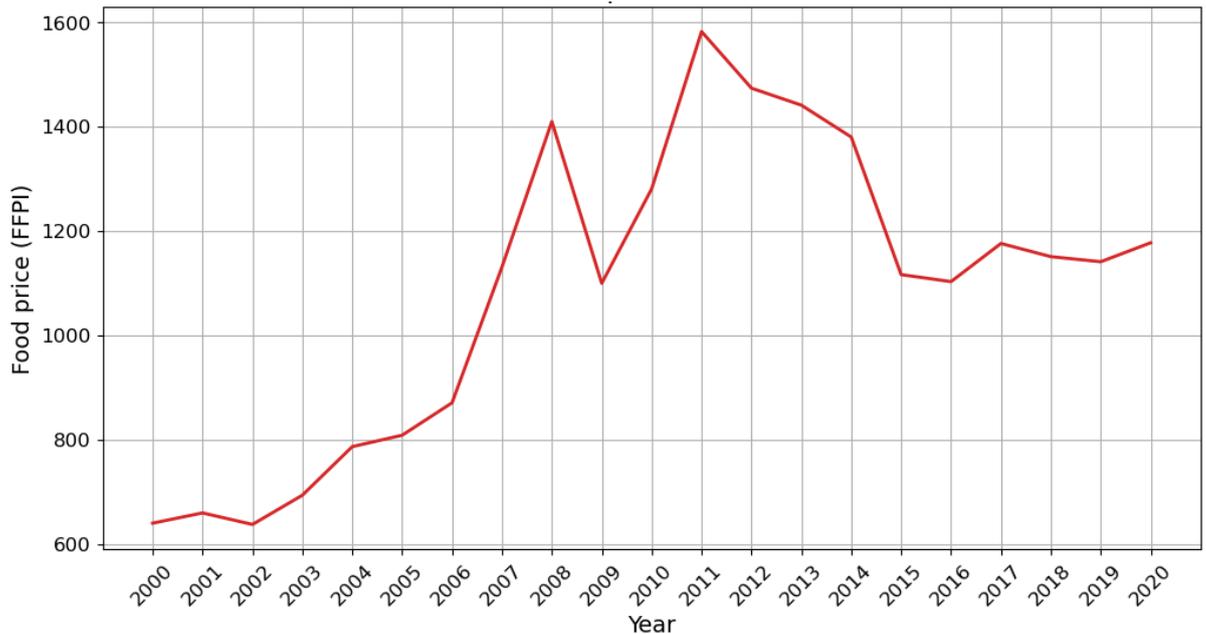

**Figure 1**. Food price (2000–2020).

*2.2 Data splitting and kernel density estimate*

The research design is shown in Fig. 2. First, multidimensional social causes derived from data mining and featuring techniques were examined. This made it possible to analyze numerous academic papers, identify popular topics, and classify them into four different categories representing different dimensions of socioeconomics. Second, machine learning was used to quantify and highlight key features. Finally, a comparative study of commonly used models was conducted to compare the fit of the data.

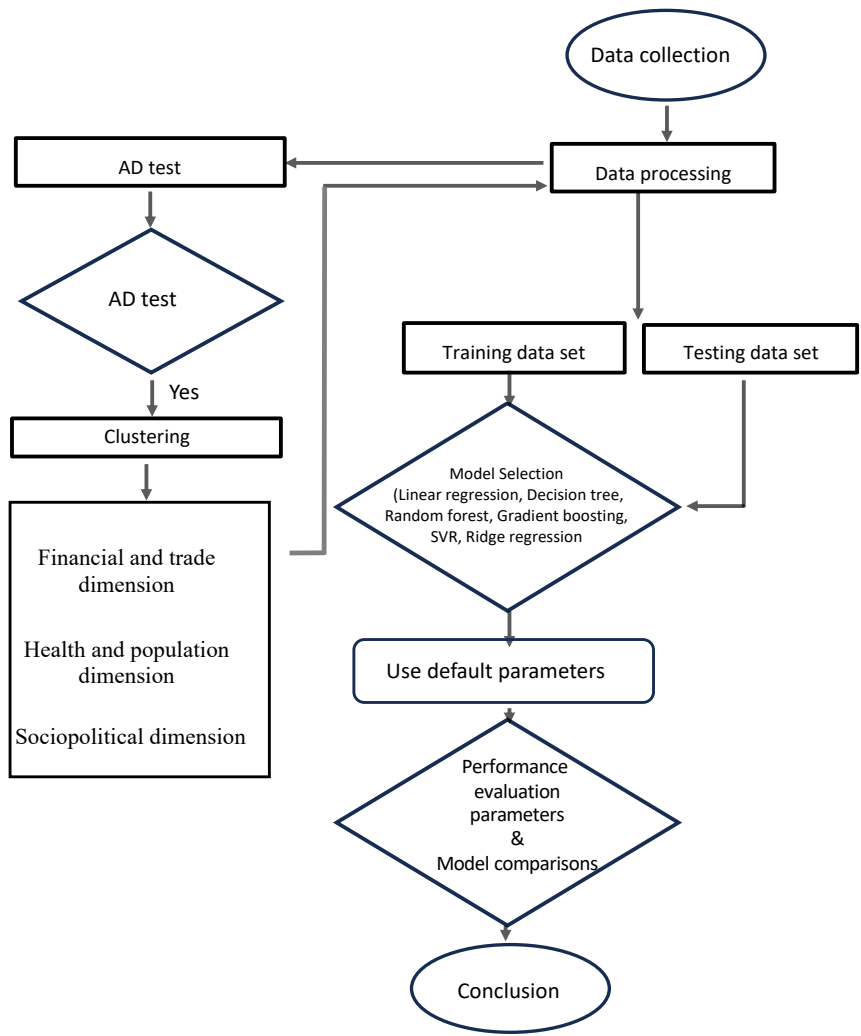

**Figure 2.** Research design

Among the data, 80% was assigned to training and 20% was assigned to testing using a fixed random seed of 42 for consistency and model reliability (Bisong, 2019; Shan, 2023). Ensuring similarity in the distribution of the training and testing sets is crucial for the model to effectively learn relevant patterns during training and generate accurate estimates during testing (Jain et al., 2000). The methodology included four main steps: the Anderson–Darling (AD) test, feature clustering, feature selection (F-value test), and identification of the top 30 key features. These steps were intended to refine the pool of features and select those that had the greatest impact on predicting food prices and safety. The AD test was performed to assess data distribution and identify normally distributed variables, as described in Supplementary Figure S1. This test was applied to each column in both the training dataset (X_train) and the testing dataset (X_test). The results were visualized using kernel density estimation plots, histograms, and fitted curves, highlighting cases with p < 0.05. In addition, the analysis code included a 'failed_tests' function to track variables that did not meet the AD testing criteria, particularly those with non-significant results (p $\geq$ 0.05), indicating a deviation from the normal distribution. Figure 3 provides examples of the variables that passed the AD test. The remaining variables of the KS exploratory data analysis are presented in Supplementary Figure S1.

This comprehensive approach ensures a thorough assessment of the distributional properties of the data. Non-normally distributed variables were transformed before further analysis. Feature clustering was then performed to group features into clusters based on their relationships and similarities, with each cluster containing highly correlated features. This approach simplified the model by reducing the number of features without losing excessive information. This was achieved by characterizing each cluster by a single feature that best represented the group's common characteristics. Feature selection using the F-value test was performed to measure the degree of linear dependence between two random variables and identify the most significant features. In this study, the F-value test helped determine the features that contributed most to the model's prediction of food prices and security. The top 30 key features in terms of impact (based on the F-value test and feature clustering) were identified from a refined pool of variables, further reducing the feature set to a manageable number without sacrificing the predictive power. These key characteristics were used in the final model to forecast food prices and security. Utilizing this systematic and robust methodological approach enables a comprehensive analysis of the variables that affect global food prices and security.

Figure 3. Examples of variables that passed the AD test.

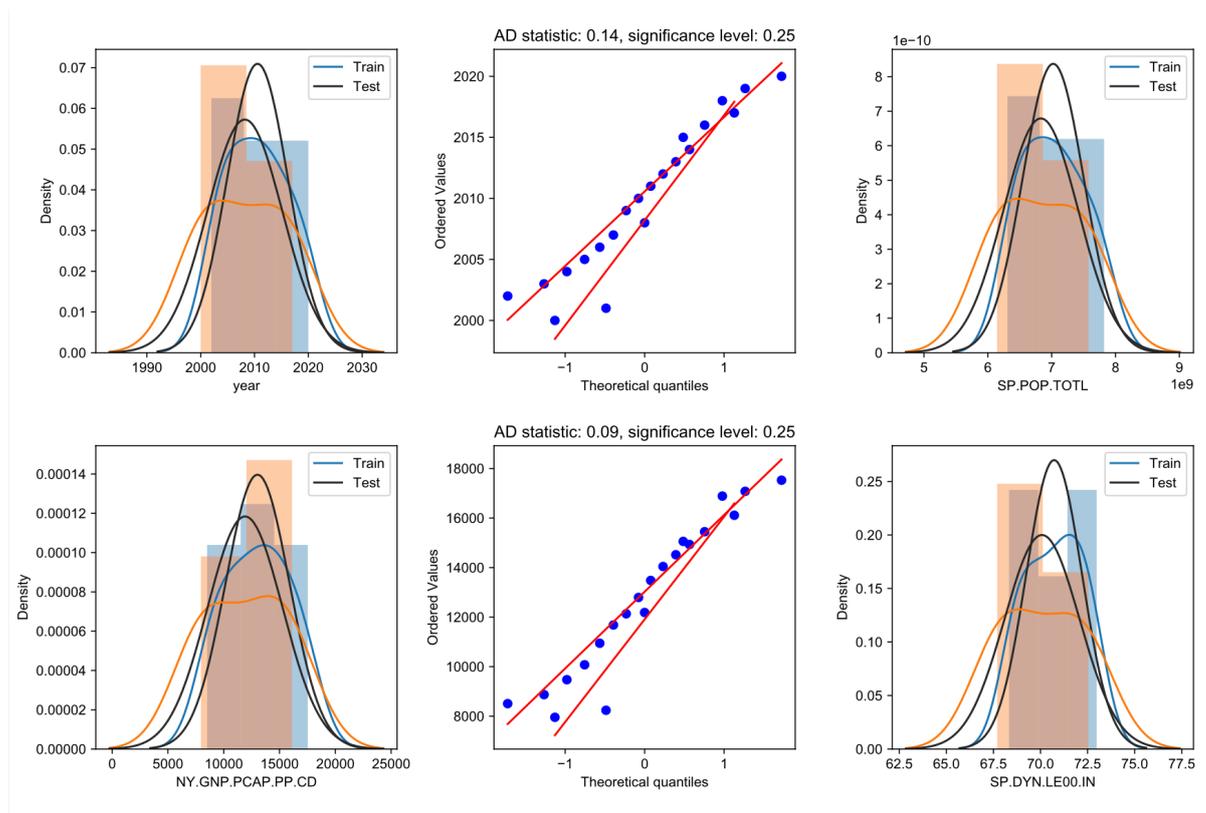

*2.3 Modeling evaluation*

A comprehensive methodology was employed by utilizing six different machine learning models (Table 1) selected for their ability to handle different aspects of the data. The models included support vector regression (SVR), ridge regression, linear regression, random forest regression, gradient boosting, and decision tree regression.

SVR is effective in high-dimensional spaces and is robust to overfitting, which is particularly useful for non-linear data (Roy and Chakraborty, 2023; Liao et al., 2024). However, the requirements for appropriate kernel and regularization parameter selection as well as computational intensity for large datasets represent well-known limitations (Ding et al., 2015).

Ridge regression can handle multicollinearity efficiently and easily. The disadvantage is that there may be poor performance on non-linear data and increased complexity due to regularization (Kigo et al., 2023).

Linear regression is simple and efficient in linear relationship modeling. However, the assumption of linear relationships and sensitivity to outliers limits its application to modeling complex relationships (Rousseeuw and Leroy, 2005; Peña and Slate, 2006).

Random forest regression creates a set of decision trees from randomly selected subsets of the training set and averages their predictions. It was chosen owing to its higher prediction accuracy and lower risk of overfitting than single decision trees (Ali et al., 2012). This approach is more computationally intensive and less interpretable than a single decision tree but provides a balance between interpretability and complexity (James et al., 2013a; Kirasich et al., 2018; Fratello and Tagliaferri, 2018).

Gradient boosting can capture complex relationships and patterns; however, while it provides insights into feature importance, it can be computationally intensive for large datasets and is prone to overfitting (Jun, 2021). It also requires tuning of hyperparameters and is difficult to interpret owing to its ensemble nature (Freeman et al., 2016; Huber et al., 2022).

Decision tree regression has good interpretability and can capture non-linear relationships. However, it is prone to overfitting and can produce overly complex trees (Lou et al., 2012; Costa and Pedreira, 2023; James et al., 2023b).

Each model was applied to ensure a robust and comprehensive analysis of the dataset, taking into account their respective strengths and weaknesses.

*Table 1. General comparison of six machine learning models used in this study*

| Machine learning model | Mechanics | Pros and cons |
|---|---|---|
| Super vector regression | Fit the best line within a threshold error margin and use different kernel functions to handle non-linear relationships. | Pros: Effective in high-dimensional spaces and with non-linear data; robust against overfitting in high-dimensional space.<br>Cons: Requires selection of appropriate kernel and regularization parameters; can be computationally intensive for large datasets. |
| Ridge regression | Minimize the sum of squared residuals with an added penalty proportional to the square of the magnitude of the coefficients. | Pros: Handles multicollinearity well; simple and computationally efficient.<br>Cons: Might not perform well with non-linear data; model complexity can be increased by introducing regularization. |
| Linear regression | Attempt to model the relationship between dependent and independent variables by fitting a linear equation to the observed data. | Pros: Simple to understand and implement; efficient for problems with linear relationships.<br>Cons: Assumes a linear relationship; sensitive to outliers; cannot model complex relationships like non-linear data. |
| Random forest regression | Create a set of decision trees from randomly selected subsets of the training set and average their predictions. | Pros: High predictive accuracy; less prone to overfitting than a single decision tree.<br>Cons: Computationally more intensive than a single decision tree; lower interpretability than that of a single decision tree but better than that of super vector regression. |
| Gradient boosting | Iteratively builds new models that focus on reducing the errors made by the previous model | Pros: Effective for capturing complex relationships and patterns; provides insights into feature importance.<br>Cons: Computationally intensive for large datasets; prone to overfitting; requires careful hyperparameter tuning; challenging to interpret owing to ensemble nature. |
| Decision tree regression | Split the dataset into subsets based on feature values; this process is recursively repeated until the tree reaches a predefined depth or purity. | Pros: High interpretability; can capture non-linear relationships.<br>Cons: Prone to overfitting, especially with complex datasets; can create overly complex trees. |

## 3. Results and Discussion

Machine learning was utilized to extract features and identify four dimensions that have a significant impact on food prices: economic aspects, social health, political and military factors, and demographic characteristics. Two methods were used to evaluate model efficiency: random forest and super vector regression. Using data mining and feature techniques, key variables related to food prices and security and their determinants were extracted from an extensive database of previous research. This process yielded 30 critical features spanning a variety of factors from carbon emission levels and population metrics to GDP per capita, military expenditure, health expenditure, and trade services.

*3.1 Shifts in food price and security in relation to complex social factors*

The relationships between significant fluctuations in food prices and security and key socioeconomic variables were investigated (Table 2; Figure S2). Changes in food prices and security were closely linked to a complex network of social factors, including economic, political, and cultural components. Understanding these relationships is crucial for developing effective strategies to manage food prices and security. The use of data mining and machine learning tools helped identify 30 key characteristics that impact food prices/security at all socioeconomic levels (Fig. 4).

*Table 2. Summary of exploratory data analysis*

| Variable | Mean | Median | Std Dev | IQR | CI Lower Bound | CI Upper Bound |
|---|---|---|---|---|---|---|
| Food Price | 1,110.02 | 1,136.14 | 285.82 | 450.58 | 969.97 | 1,250.06 |
| NY.GNP.ATLS.CD (×10^12) | 64.89 | 67.65 | 18.00 | 27.73 | 56.07 | 73.71 |
| NY.GNP.PCAP.CD | 9,116.49 | 9,646.94 | 1,993.31 | 3,149.37 | 8,139.79 | 10,093.2 |
| NY.GDP.MKTP.CD (×10^12) | 65.38 | 70.25 | 17.55 | 27.33 | 56.78 | 73.97 |
| NV.AGR.TOTL.ZS | 3.76 | 3.92 | 0.39 | 0.70 | 3.57 | 3.96 |
| NE.EXP.GNFS.ZS | 27.90 | 28.52 | 2.26 | 3.47 | 26.79 | 29.01 |
| NE.IMP.GNFS.ZS | 27.29 | 27.85 | 2.06 | 3.41 | 26.28 | 28.30 |
| TG.VAL.TOTL.GD.ZS | 44.90 | 45.20 | 3.71 | 5.78 | 43.08 | 46.72 |

| Variable | Mean | Median | Std Dev | IQR | CI Lower Bound | CI Upper Bound |
|---|---|---|---|---|---|---|
| BM.TRF.PWKR.CD.DT ($\times 10^{11}$) | 3.12 | 3.33 | 1.09 | 1.93 | 2.59 | 3.66 |
| DT.ODA.ODAT.PC.ZS | 17.96 | 18.55 | 4.26 | 5.17 | 15.87 | 20.05 |
| NE.RSB.GNFS.ZS | 0.61 | 0.65 | 0.25 | 0.32 | 0.49 | 0.74 |
| NY.GDS.TOTL.ZS | 25.55 | 26.08 | 1.41 | 2.09 | 24.85 | 26.24 |
| NE.CON.PRVT.ZS | 57.49 | 57.04 | 1.46 | 1.87 | 56.78 | 58.21 |
| NY.GDP.MINR.RT.ZS | 0.33 | 0.29 | 0.21 | 0.24 | 0.23 | 0.43 |
| BX.TRF.PWKR.DT.GD.ZS | 0.64 | 0.65 | 0.10 | 0.15 | 0.59 | 0.69 |
| MS.MIL.XPND.CD ($\times 10^{12}$) | 1.50 | 1.65 | 0.36 | 0.56 | 1.33 | 1.68 |
| NY.GDP.PCAP.CD | 9,188.97 | 9,881.72 | 1,933.54 | 3,095.88 | 8,241.55 | 10,136.39 |
| SH.DYN.MORT | 52.46 | 50.45 | 10.86 | 17.38 | 47.14 | 57.79 |
| SP.DYN.LE00.FE.IN | 73.23 | 73.36 | 1.54 | 2.51 | 72.48 | 73.99 |
| SH.MMR.RISK.ZS | 0.62 | 0.60 | 0.10 | 0.17 | 0.57 | 0.67 |
| EN.URB.LCTY.UR.ZS | 16.15 | 16.07 | 0.18 | 0.26 | 16.06 | 16.24 |
| SI.POV.GAPS | 4.89 | 4.35 | 2.07 | 3.30 | 3.87 | 5.90 |
| SI.POV.LMIC.GP | 13.83 | 13.25 | 4.42 | 7.03 | 11.66 | 15.99 |
| SI.POV.DDAY | 16.23 | 15.25 | 6.24 | 10.07 | 13.17 | 19.28 |
| SH.IMM.MEAS | 81.44 | 83.91 | 4.58 | 6.17 | 79.19 | 83.68 |
| SE.PRM.CMPT.ZS | 87.74 | 88.98 | 2.28 | 3.48 | 86.63 | 88.86 |
| SE.SEC.ENRR | 70.21 | 71.67 | 5.59 | 10.44 | 67.47 | 72.95 |
| SE.ENR.PRSC.FM.ZS | 0.97 | 0.97 | 0.02 | 0.03 | 0.96 | 0.98 |
| EN.ATM.CO2E.PC | 4.47 | 4.54 | 0.22 | 0.27 | 4.36 | 4.58 |
| IT.CEL.SETS.P2 | 69.35 | 79.59 | 32.16 | 56.54 | 53.59 | 85.11 |
| SE.XPD.TOTL.GD.ZS | 4.17 | 4.20 | 0.18 | 0.25 | 4.09 | 4.26 |

*This table presents a detailed summary of the exploratory data analysis, showcasing the central tendency (Mean, Median), variability (Std Dev, IQR), and 95% confidence interval bounds (CI Lower Bound, CI Upper Bound) for each variable in the subselection of the database. Hereby, "×10^n" denotes the magnitude for large numbers and precision for each numerical value to maintain clarity and readability.*

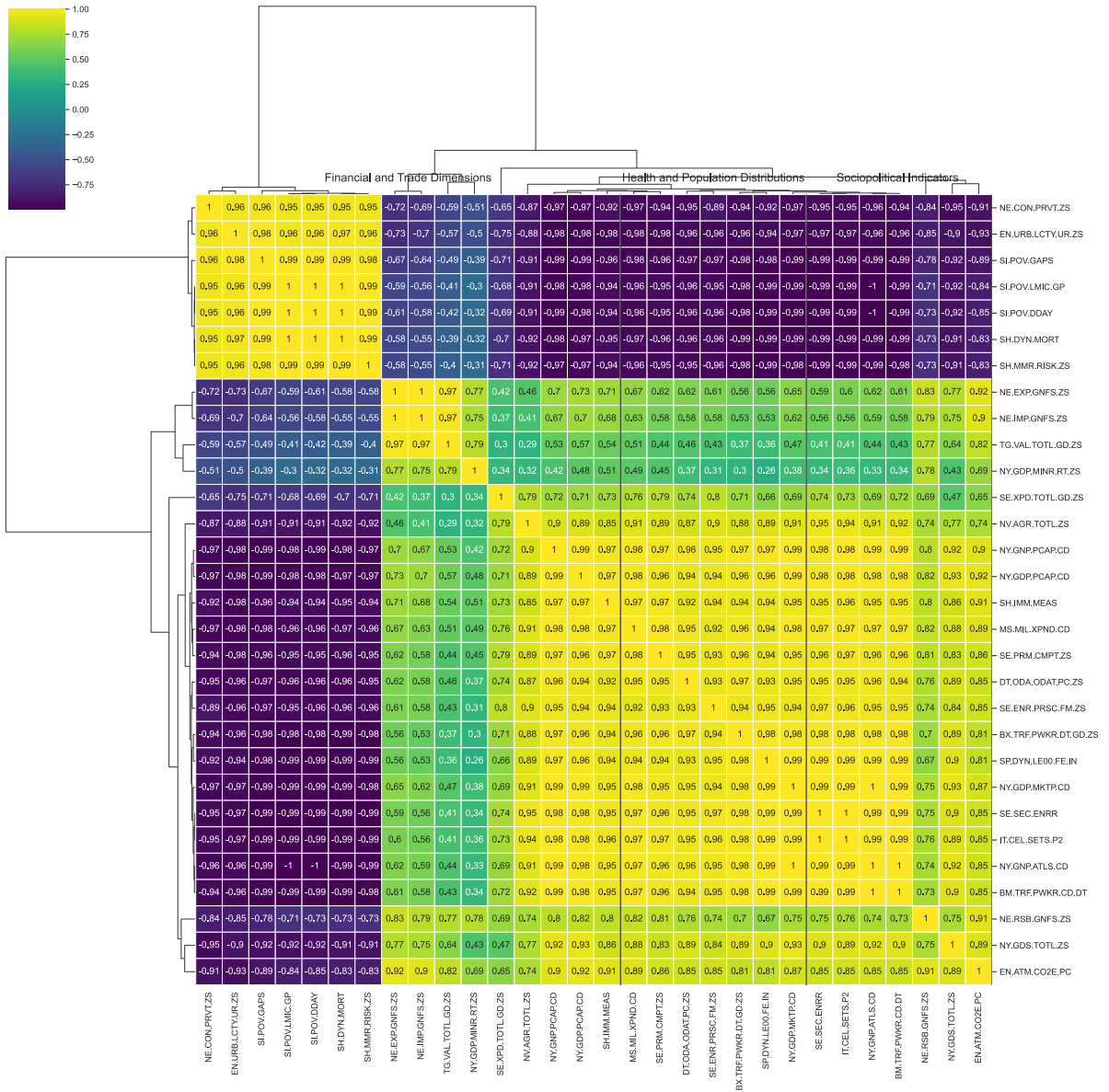

**Figure 4.** Heatmap of clustering results representing an interconnected variable matrix. The cell at the intersection of row i and column j shows the correlation between the i_th and j_th features. Contrasting colors indicate the strength and direction of the correlation; a value close to 1 indicates a strong positive correlation, a value close to −1 indicates a strong negative correlation, and a value close to 0 indicates no linear relationship. The clustering algorithm organizes rows and columns to group similar features together, improving interpretability.

3.1.1 Economic indicators and impacts on global food prices

Economic indicators play a central role in the complex network that shapes global food price security (Table 3). Gross national income (GNI), both at the state and per capita levels ("NY.GNP.ATLS.CD" and "NY.GNP.PCAP.CD," respectively), serves as an important measure of economic performance. Although a higher GNI often allows countries to invest more in agricultural development and food imports, it does not necessarily ensure equitable food security, especially without an equal distribution of wealth (Smith et al., 2012). Furthermore, higher GNI per capita can increase individual purchasing power but also stimulate demand, potentially leading to price inflation. GDP and its per capita variant ("NY.GDP.MKTP.CD" and "NY.GDP.PCAP.CD") are also critical indicators. A high GDP is typically indicative of enhanced food security, as it allows investment in agricultural production (Nelson et al., 2010; Wheeler and Von Braun, 2013). However, interpreting GDP alone can be misleading because the percentage of GDP that comes from agriculture, forestry, and fishing ("NV.AGR.TOTL.ZS") can provide more insight into the stability of a country's food supply. A higher percentage of agriculture often signals stability but can expose the country to global price shocks, especially if it is too dependent on a single sector.

Trade dynamics, including export and import rates ("NE.EXP.GNFS.ZS" and "NE.IMP.GNFS.ZS") and the total value of goods trade ("TG.VAL.TOTL.GD.ZS"), exert a direct influence on food prices. For example, countries with high export rates, especially those in the agricultural sector, can deplete local food supplies and increase domestic prices. Conversely, dependence on food imports exposes a country to international price fluctuations (Barrett, 2010). Financial inflows and outflows, such as personal remittances both paid and received ("BM.TRF.PWKR.CD.DT" and "BX.TRF.PWKR.DT.GD.ZS"), can influence household income (i.e., food security). However, higher income can also increase demand and, therefore, food prices. Official development assistance ("DT.ODA.ODAT.PC.ZS"), often directed at agricultural and food security programs, can stabilize food prices in recipient countries, but it can also create a negative balance of goods and services ("NE.RSB.GNFS.ZS"), which make them vulnerable to international price fluctuations (Prakash, 2011).

Economic factors, such as domestic savings ("NY.GDS.TOTL.ZS") and private consumption rates ("NE.CON.PRVT.ZS") can also influence food prices. Increased domestic savings can enable better investment opportunities in agriculture and potentially stabilize local food prices. In contrast, increased private consumption can drive up demand and, thus, food prices.

Military spending ("MS.MIL.XPND.CD") and the value added by the mining sector ("NY.GDP.MINR.RT.ZS") can also have an indirect impact on food security. For example. countries with high military spending may invest less in agriculture, creating an imbalance that could impact food security (Smith et al., 2012).

Understanding these diverse economic indicators is critical to any comprehensive strategy for managing global food price volatility. An effective policy response requires an integrated approach that includes these economic indicators as well as social, political, and demographic variables.

*Table 3. Economic indicators impacting global food prices*

| Indicator | Interpretation |
| --- | --- |
| NY.GNP.ATLS.CD | Gross national income (GNI) calculated using the Atlas method in current U.S. dollars |
| NY.GNP.PCAP.CD | GNI per capita, computed using the Atlas method in current U.S. dollars |
| NY.GDP.MKTP.CD | GDP in current U.S. dollars |
| NV.AGR.TOTL.ZS | Percentage of GDP generated from agriculture, forestry, and fishing |
| NE.EXP.GNFS.ZS | Exports of goods and services as a percentage of GDP |
| NE.IMP.GNFS.ZS | Imports of goods and services as a percentage of GDP |
| TG.VAL.TOTL.GD.ZS | Merchandise trade as a percentage of GDP |
| BM.TRF.PWKR.CD.DT | Personal remittances paid in current U.S. dollars |
| DT.ODA.ODAT.PC.ZS | Net official development assistance (ODA) received per capita in current U.S. dollars |
| NE.RSB.GNFS.ZS | Net balance of goods and services as a percentage of GDP. |
| NY.GDS.TOTL.ZS | Gross domestic savings as a percentage of GDP |
| NE.CON.PRVT.ZS | Private consumption as a percentage of GDP |
| NY.GDP.MINR.RT.ZS | Value added by the mining sector as a percentage of GDP |
| BX.TRF.PWKR.DT.GD.ZS | Personal remittances received as a percentage of GDP |
| MS.MIL.XPND.CD | Military expenditure in current U.S. dollars |
| NY.GDP.PCAP.CD | GDP per capita in current U.S. dollars |

3.1.2 Demographic indicators and impacts on global food prices

Demographic indicators provide another invaluable basis for understanding the complexities of global food prices (Table 4). Mortality rates for children under 5 years of age ("SH.DYN.MORT"), life expectancy for women ("SP.DYN.LE00.FE.IN"), and maternal mortality risks ("SH.MMR.RISK.ZS") serve as important measures of societal well-being and indirectly reflect the state of food and nutritional security (Barrett, 2010). Higher child and maternal mortality rates often indicate underlying problems related to food insecurity and malnutrition, which can impact food price structures. In addition, the demographic compositions of urban and rural populations ("EN.URB.LCTY.UR.ZS") have an important influence on food distribution systems. Urban areas tend to rely on complicated logistics and often rely on imported food, making them vulnerable to global price fluctuations, which in turn exacerbates food security problems (Smith et al., 2012). Poverty-related indicators, such as the poverty gap at $1.90 a day ("SI.POV.GAPS") and poverty headcount ratios at low–middle-income levels ("SI.POV.LMIC.GP") and at $1.90/d ("SI.POV.DDAY"), explain socioeconomic differences within a population. These indicators are particularly relevant because those living near or below the poverty line are more vulnerable to small changes in food prices. Their reduced purchasing power is related to their ability to access nutritious food, which negatively impacts the stability and security of food prices (Barrett, 2010; Smith et al., 2012). The complexity of these demographic indicators highlights the need for an integrated policymaking approach. Considering demographic factors along with economic, social, and political variables provides a comprehensive strategy for managing the global food price ecosystem.

*Table 4. Demographic indicators affecting global food prices*

| Indicator | Interpretation |
|---|---|
| SH.DYN.MORT | Mortality rate for children under 5 years, expressed per 1,000 live births |
| SP.DYN.LE00.FE.IN | Life expectancy at birth for females |
| SH.MMR.RISK.ZS | Maternal mortality risk |
| EN.URB.LCTY.UR.ZS | Percentage of the population living in urban areas |
| SI.POV.GAPS | Poverty gap at $1.90/d (2011 international prices) as a percentage. |
| SI.POV.LMIC.GP | Poverty headcount ratio at low–middle-income levels |
| SI.POV.DDAY | Poverty headcount ratio at $1.90/d (2011 international prices) as a percentage of the population |

3.1.3 Sociopolitical features and impacts on global food prices

Among sociopolitical features (Table 5), the percentage of children vaccinated against measles ("SH.IMM.MEAS") serves as an index of public health infrastructure, which in turn, impacts agricultural labor productivity (Barrett, 2010). Likewise, primary school attainment ("SE.PRM.CMPT.ZS") and gross secondary school enrollment rates ("SE.SEC.ENRR") reflect a country's investment in human capital, which can influence the skills available for agricultural productivity and innovation, with a long-term impact on food prices (Smith et al., 2012). The gender parity index for school enrollment in primary and secondary education ("SE.ENR.PRSC.FM.ZS") can also indicate the level of gender equality in a society, which has been shown to influence food security through more equitable distribution and decision-making within households and communities (Barrett, 2010). $CO_2$ emissions per capita ("EN.ATM.CO2E.PC") are often used to indicate the level of industrialization of a country and can potentially impact the sustainability of agriculture and, therefore, food prices. Mobile subscriptions per 100 people ("IT.CEL.SETS.P2") serve as a measure of the rate of technological dissemination, which can impact food prices through market efficiency and information symmetry (Smith et al., 2012). Lastly, the total expenditure on education as a percentage of GDP ("SE.XPD.TOTL.GD.ZS") represents a commitment to skill development that could improve agricultural processes and thereby promote food price stability (Barrett, 2010). The data suggest that investments in education, gender equity, technology adoption, and environmental sustainability jointly influence global food prices and global food security.

The multidimensional influences of these sociopolitical indicators on food price security highlight the need for an integrated policy approach. They also highlight complex connections between education, gender equity, technological reach, and environmental sustainability, and their collective impact on food prices and security. These categories play different but interrelated roles. Economic indicators mainly refer to financial and trade dimensions that directly affect food security. Demographic indicators provide insights into health and population distributions that are essential for targeted food security interventions. Lastly, sociopolitical indicators reflect a society's level of development, infrastructure, and values, which indirectly influence food security through policies and awareness.

*Table 5. Sociopolitical factors affecting global food prices*

| Indicator | Interpretation |
|---|---|
| SH.IMM.MEAS | Percentage of children between the ages of 12–23 months who have been immunized against measles |
| SE.PRM.CMPT.ZS | Primary school completion rate, measured as a percentage of the relevant age group |
| SE.SEC.ENRR | Gross enrollment rate in secondary education |
| SE.ENR.PRSC.FM.ZS | Gender parity index for school enrollment in primary and secondary education |
| EN.ATM.CO2E.PC | $CO_2$ emissions per capita in metric tons |
| IT.CEL.SETS.P2 | Mobile cellular subscriptions per 100 people |
| SE.XPD.TOTL.GD.ZS | Total expenditure on education as a percentage of GDP |

*3.2 Model efficacy based on SVR*

Ridge regression and SVR have shown excellent performance in predicting changes in food prices influenced by various socioeconomic factors, as detailed in Table 6. These machine learning algorithms excel at accurately forecasting food prices and security, as well as deciphering the intricate connections among diverse social factors. However, ridge regression faces challenges related to its management of high-dimensional spaces and the extensive number of training examples. Determining the optimal penalty value for ridge regression requires not only a deep understanding of the algorithm but also considerable experimentation and validation efforts. This is crucial for achieving a model that is both accurate and generalizable. In contrast, SVR distinguishes itself through its resistance to overfitting and its proficiency in handling non-linear relationships.

*Table 6. Model performance*

| Model | MAE | RMSE | $R^2$ |
|---|---|---|---|
| SVR | 0.471065 | 0.315638 | 0.993144 |
| Ridge regression | 0.768931 | 1.123438 | 0.975596 |
| Linear regression | 1.046580 | 2.525464 | 0.945140 |
| Gradient boosting | 2.298773 | 9.382184 | 0.796194 |
| Random forest | 2.655952 | 11.999760 | 0.739333 |
| Decision tree | 3.792857 | 22.349643 | 0.514507 |

MAE: mean absolute error, RMSE: root mean squared error.

**Algorithm 1: Support vector regression (SVR) for food price impact**

**Input:**

- Feature matrix **X** with economic, demographic, and environmental indicators.
- Target vector $y$ representing global food prices.

Output:

- Optimized SVR model parameters.
- Evaluation metrics: $MAE = 0.471065, MSE = 0.315638, R^2 = 0.993144$.

SVR Function:

- $f(\mathbf{x}) = \mathbf{w}^\top \phi(\mathbf{x}) + b$
- Subject to the optimization of:
  - $\min_{\mathbf{w},b,\xi,\xi^*} \frac{1}{2} \parallel \mathbf{w} \parallel^2 + C \sum_{i=1}^{n}(\xi_i + \xi_i^*)$
  - Constraints for $\xi_i, \xi_i^*$ given $\epsilon$ insensitivity zone:
    - $y_i - \mathbf{w}^\top \phi(\mathbf{x}_i) - b \leq \epsilon + \xi_i$
    - $\mathbf{w}^\top \phi(\mathbf{x}_i) + b - y_i \leq \epsilon + \xi_i^*$
    - $\xi_i, \xi_i^* \geq 0$

Procedure:

1. Preprocess **X** by scaling features to a mean of zero and a variance of one.
2. Initialize the SVR model with hyperparameter space comprising:
   - Regularization parameter $C$,
   - Epsilon $\epsilon$,
   - Kernel coefficient $\gamma$,
   - Kernel type kernel.

3. Apply 5-fold cross-validation grid search to determine the optimal set of parameters for the SVR model.

4. Train the SVR model using the optimal parameters on the scaled dataset.

5. Predict food prices using the trained SVR model, employing the kernel trick $K(\mathbf{x}_i, \mathbf{x}_j) = \phi(\mathbf{x}_i)^\top \phi(\mathbf{x}_j)$ to project data into a higher-dimensional space[2].

6. Compute performance metrics: MAE, MSE, and $R^2$ using the SVR model's predictions.

**End Procedure**

*3.3 Machine learning reveals the importance of interdisciplinary approaches in addressing food price security*

To study food price certainty, advanced machine learning techniques were used, including ridge regression, decision tree regressors, random forest regression, and SVR, which demonstrated superior predictive performance. Through data mining, feature extraction, and automated database interactions, a refined list of 30 key multidimensional variables spanning economic, social, and political factors was identified through a comprehensive review of existing research. These variables provide a basis for understanding the complex dynamics that influence food prices and, therefore, security. The high effectiveness of SVR as well as the commendable performance of linear and ridge regression models support the potential of machine learning as a tool in this context.

The application of machine learning techniques not only improved the predictive power but also highlighted the essential role of interdisciplinary approaches that enable the identification and integration of critical socioeconomic factors, such as health expenditure and economic contributions, which are crucial for formulating robust strategies to address the diverse challenges. The results highlight the need for an interdisciplinary approach that incorporates expertise from relevant fields (e.g., economics, political science, environmental science, and public health) to address the complexities of food price security at a global level. For example, understanding the political stability index in the context of healthcare spending can provide nuanced perspectives that may be missed with a single-discipline focus. Likewise, considering environmental variables alongside economic indicators, such as GDP or GNI, allows for a more informed analysis of food production and distribution networks.

---

[2] In this algorithm, $\phi(\mathbf{x})$ represents feature mapping to higher dimensional space, which is implicit in the kernel trick with $K$ being the kernel function. The regularization parameter $C$ controls the trade-off between the model's complexity and the degree to which deviations larger than $\epsilon$ are tolerated. $\xi$ and $\xi^*$ are slack variables that allow for violations of the $\epsilon$ insensitivity zone, which is essential for capturing errors in the model. This optimization is typically solved using a quadratic programming solver in the dual space, which is reflected in the grid search step of the algorithm.

Such an interdisciplinary perspective is invaluable for policymakers who often operate at the intersection of these different but interconnected areas. The insights gained from machine learning models provide a quantifiable basis for targeted policy interventions. They provide empirical evidence for strategies to stabilize food prices, thereby improving food security and the general well-being of people.

*3.4 Study limitations and prospects*

This study used macro-level data rather than individual factors for its analysis. Given this approach and the methodology used to address macro-level issues, there are inherent limitations in tackling specific factor issues. A major limitation is the data range of the study, which extends from 2000 to 2022. This timeframe may not adequately represent long-term trends affecting global food price security. Furthermore, the reliance on data from the WDI and the FAO diminishes the scope of the study and may miss crucial factors not included in these datasets. Another limitation is the predominantly quantitative focus of the study, which may not fully take into account certain qualitative aspects such as political events and cultural changes. While the study effectively presents important factors and their associations with food prices, it is important to note that this does not necessarily imply a causal relationship.

Establishing causal relationships in multilevel data remains a challenging task, especially given current methodological limitations.

The ever-evolving landscape of global food prices pens several avenues for future research that deserve close attention. One is to tailor models to specific regions, considering geographical differences in socioeconomic, political, and environmental conditions. This adaptation is crucial for the accurate formulation and implementation of policies in different contexts. In addition, there is an increasing need to investigate the effectiveness of new computational methods, including advanced machine learning and artificial intelligence algorithms, for predicting food prices and security. These innovative techniques can potentially capture the complexity of a problem more effectively than traditional statistical models. In addition to methodological richness, longitudinal studies that examine changes in food prices and security over longer periods can provide invaluable insight into evolving trends and challenges, enabling more accurate forecasts for future scenarios. While this study represents progress in identifying characteristics that impact food prices and security, subsequent studies could benefit from including additional variables, such as impacts of climate change or international trade relations, thereby enriching both the analytical framework and policy recommendations.

From a governance perspective, integrating insights from policy studies into future research can help develop robust strategies for managing food price volatility and security issues. This policy-oriented approach provides a structured framework for decision makers and stakeholders concerned with food security. Additionally, assessing the generalizability of the results in different local and global settings will refine the models for broader applicability and real-world impacts. Finally, given the complexity of food security challenges, future research should promote interdisciplinary collaboration between fields such as economics, social sciences, environmental science, and computer science.

Interdisciplinary engagement is essential to promote innovative solutions to complex and pressing global problems. Future research in these areas can build on the methodological framework presented herein to contribute not only to academic discourse but also to pragmatic data-driven strategies to address this important problem.

**4. Conclusion**

This study highlights the importance of economic and demographic indicators for global food price security. The results showed that factors such as GNI and GDP are indicators of more than just a country's economic health; they are essential for understanding and ensuring food availability for the population. These economic indicators, coupled with demographic metrics, such as child mortality rates and poverty levels, provide a multifaceted view of food security, which is closely linked to household income and urbanization.

The developed approach including data mining and machine learning techniques identified 30 key characteristics that influence food security across multiple socioeconomic dimensions. The use of diverse models, including SVR, ridge regression, and decision tree analysis, provided new insights into the factors that influence food prices and security. The effectiveness of these models highlights the need for an interdisciplinary strategy to address the complexities of food price security.

In summary, addressing global food price security is a multifaceted challenge that requires a comprehensive and multidimensional approach. As well as economic considerations, this requires a deep understanding of demographic characteristics. Both sets of indicators are critical to developing effective policy responses that address the complexities of global food price security. The results showed significant shifts in the food price index, particularly related to military expenditure, health care expenditure, and economic contributions. This study advanced our understanding of the complex relationship between socioeconomic variables and food price security. By utilizing machine learning techniques and considering multiple dimensions, policymakers can make informed decisions to improve food prices and food security on a global scale. In addition, the results may be generalizable to other situations if further study and local adaptations are made.

**Acknowledgments**

Not applicable.

**Funding**

This research received no specific grant from any funding agency in the public, commercial, or not-for-profit sectors.

**Competing interests**

The author declares that there are no competing interests to report.

**Author contributions**

As the sole author, SS independently conducted the entire research process and preparation of the manuscript.

**Data Availability**

The data that support the findings of this study are available at

https://github.com/shanshanfy/GlobalFoodPrices.git

**Code Availability**

The analysis code can be accessed at https://github.com/shanshanfy/GlobalFoodPrices.git

**Developer Environment Availability**

The developer environment 'GlobalFoodPricesPackages.yaml' file for the Conda open-source package management system is provided through: https://github.com/shanshanfy/GlobalFoodPrices.git. This file allows for isolated environments to manage packages without interference. The file contains the configuration of the project's Python environment, including channels, dependencies, and library versions.

# Supplementary Materials

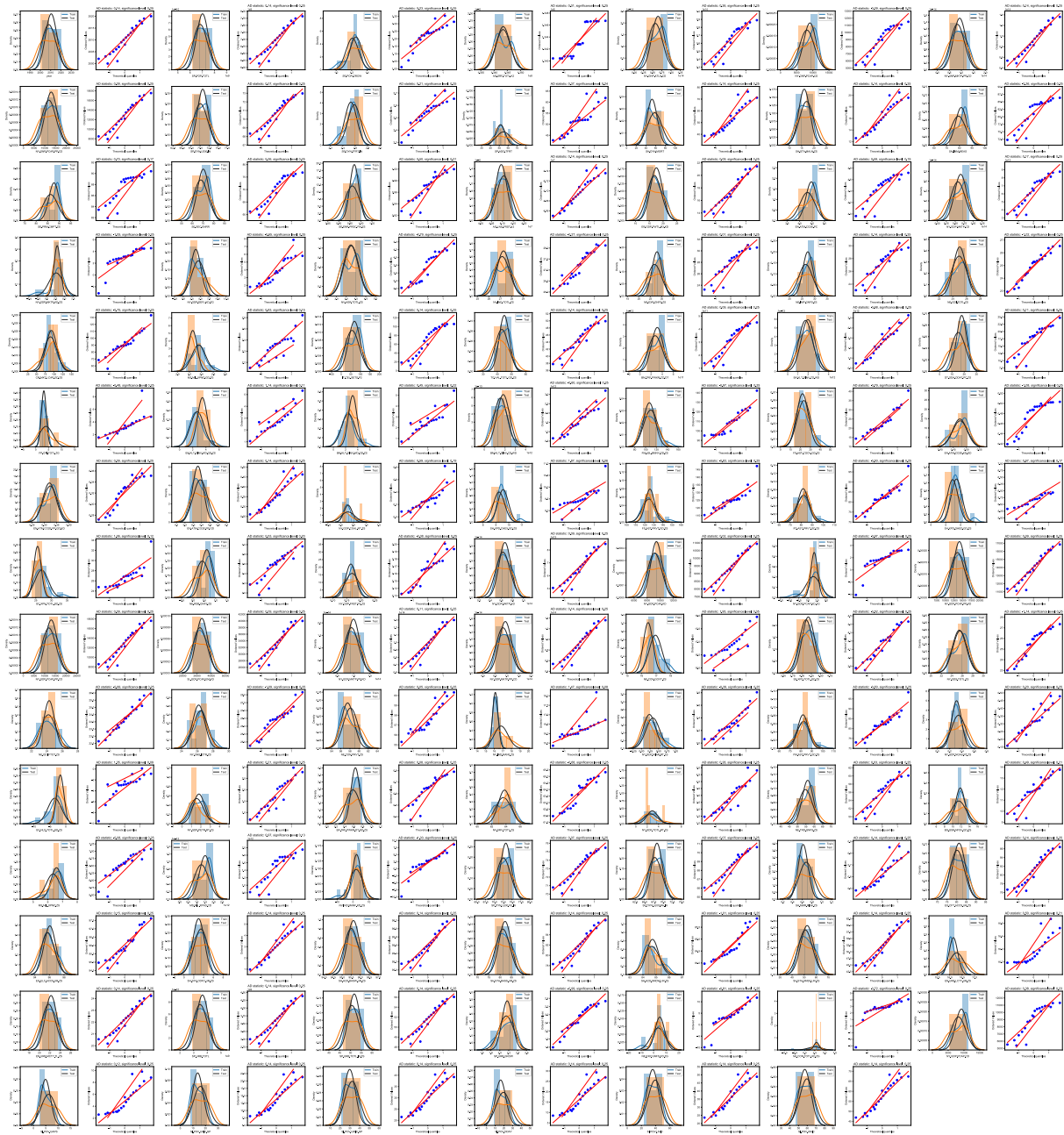

**Figure S1.** Variables that passed the Anderson–Darling (AD) test.

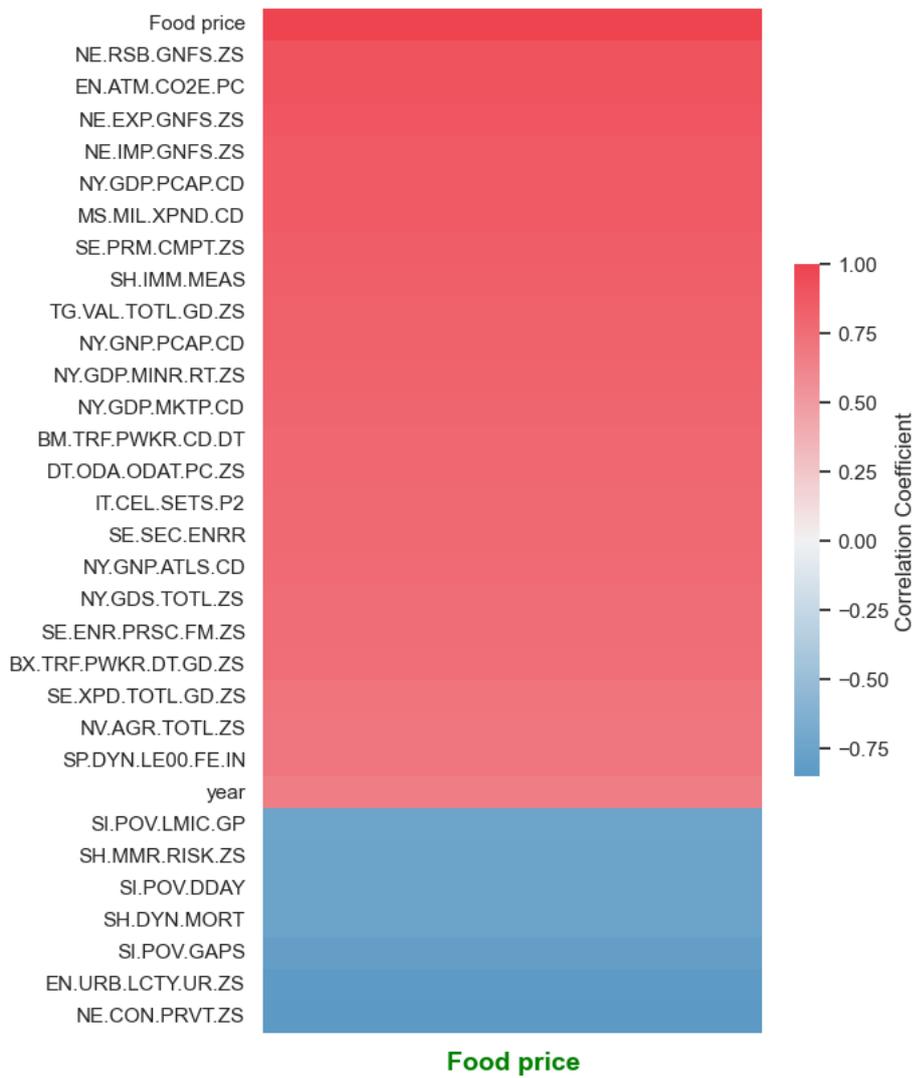

**Figure S2.** Exploratory data analysis of data behavior showing each key feature's correlation with food price and security.

Table S1. The target and 104 features.

| Abbreviation | Full Name |
| --- | --- |
| year | Year |
| FFPI | FAO Food Price Index |
| SP.POP.TOTL | Total Population |
| SP.POP.GROW | Population Growth (annual %) |

| Abbreviation | Full Name |
| --- | --- |
| AG.SRF.TOTL.K2 | Total Surface Area (sq. km) |
| NY.GNP.ATLS.CD | GNP, Atlas method (current US$) |
| NY.GNP.PCAP.CD | GNP per capita, Atlas method (current US$) |
| NY.GNP.MKTP.PP.CD | GNP, PPP (current international $) |
| NY.GNP.PCAP.PP.CD | GNP per capita, PPP (current international $) |
| SP.DYN.LE00.IN | Life Expectancy at Birth, total (years) |
| SP.DYN.TFRT.IN | Total Fertility Rate (births per woman) |
| SP.ADO.TFRT | Adolescent Fertility Rate (births per 1,000 women ages 15-19) |
| SH.DYN.MORT | Under-5 Mortality Rate (per 1,000 live births) |
| SH.STA.MALN.ZS | Prevalence of Undernourishment (%) |
| SH.IMM.MEAS | Immunization, Measles (% of children ages 12-23 months) |
| SE.PRM.CMPT.ZS | Primary Completion Rate (%) |
| SE.SEC.ENRR | Secondary Enrollment Ratio (%) |
| SE.ENR.PRSC.FM.ZS | School Enrollment, Primary (female to male ratio) |
| AG.LND.FRST.K2 | Forest Area (sq. km) |
| ER.GDP.FWTL.M3.KD | Freshwater withdrawals as a proportion of GDP |
| EN.ATM.CO2E.PC | $CO_2$ Emissions (metric tons per capita) |
| NY.GDP.MKTP.CD | GDP at market prices (current US$) |
| NY.GDP.MKTP.KD.ZG | GDP Growth (annual %) |
| NY.GDP.DEFL.KD.ZG | GDP Deflator (annual %) |

| Abbreviation | Full Name |
| --- | --- |
| NV.AGR.TOTL.ZS | Agriculture, value added (% of GDP) |
| NV.IND.TOTL.ZS | Industry, value added (% of GDP) |
| NE.EXP.GNFS.ZS | Exports of Goods and Services (% of GDP) |
| NE.IMP.GNFS.ZS | Imports of Goods and Services (% of GDP) |
| NE.GDI.TOTL.ZS | Gross Domestic Investment (% of GDP) |
| CM.MKT.LCAP.GD.ZS | Market capitalization of listed companies (% of GDP) |
| MS.MIL.XPND.GD.ZS | Military Expenditure (% of GDP) |
| IT.CEL.SETS.P2 | Mobile Cellular Subscriptions (per 100 people) |
| TG.VAL.TOTL.GD.ZS | Trade (% of GDP) |
| BM.TRF.PWKR.CD.DT | Personal Remittances, Received (current US$) |
| BX.KLT.DINV.CD.WD | Foreign Direct Investment, Net Inflows (BoP, current US$) |
| DT.ODA.ODAT.PC.ZS | Official Development Assistance and Official Aid Received (per capita) |
| FP.CPI.TOTL.ZG | Inflation, Consumer Prices (annual %) |
| BX.KLT.DINV.WD.GD.ZS | FDI, Net Inflows (% of GDP) |
| BM.KLT.DINV.WD.GD.ZS | FDI, Net Outflows (% of GDP) |
| BM.KLT.DINV.CD.WD | Foreign Direct Investment, Net Outflows (BoP, current US$) |
| FM.LBL.BMNY.GD.ZS | Broad Money (% of GDP) |
| FS.AST.CGOV.GD.ZS | Claims on Central Government (% of GDP) |
| EN.ATM.CO2E.KD.GD | $CO_2$ Emissions (kg per 2010 US$ of GDP) |
| EN.ATM.CO2E.PP.GD.KD | $CO_2$ Emissions (kg per PPP $ of GDP) |

| Abbreviation | Full Name |
|---|---|
| EN.ATM.CO2E.PP.GD | $CO_2$ Emissions (metric tons per capita and PPP of GDP) |
| NY.GDP.COAL.RT.ZS | Coal Rent (% of GDP) |
| SH.XPD.CHEX.GD.ZS | Current Health Expenditure (% of GDP) |
| FS.AST.PRVT.GD.ZS | Domestic Credit to Private Sector (% of GDP) |
| FD.AST.PRVT.GD.ZS | Domestic Credit to Private Sector by Banks (% of GDP) |
| SH.XPD.GHED.GD.ZS | Government Health Expenditure (% of GDP) |
| GC.XPN.TOTL.GD.ZS | Government Expenditure (% of GDP) |
| NE.RSB.GNFS.ZS | Reserves of Foreign Exchange and Gold (% of GDP) |
| NY.GDP.FRST.RT.ZS | Forest Rent (% of GDP) |
| NY.GDP.MKTP.KD | GDP at market prices (constant 2010 US$) |
| NY.GDP.PCAP.KD | GDP per capita (constant 2010 US$) |
| NY.GDP.PCAP.KD.ZG | GDP per capita growth (annual %) |
| NY.GDP.PCAP.PP.KD | GDP per capita, PPP (constant 2011 international $) |
| NY.GDP.PCAP.PP.CD | GDP per capita, PPP (current international $) |
| SL.GDP.PCAP.EM.KD | GDP per capita, employed (constant 2010 US$) |
| NY.GDP.MKTP.PP.KD | GDP, PPP (constant 2011 international $) |
| NY.GDP.MKTP.PP.CD | GDP, PPP (current international $) |
| NE.CON.GOVT.ZS | Government Consumption (% of GDP) |
| SE.XPD.TOTL.GD.ZS | Total Expenditure on Education (% of GDP) |
| NY.GDS.TOTL.ZS | Gross Domestic Savings (% of GDP) |

| Abbreviation | Full Name |
| --- | --- |
| NE.GDI.FTOT.ZS | Gross Fixed Capital Formation (% of GDP) |
| NY.GNS.ICTR.ZS | Gross National Savings (% of GDP) |
| NE.CON.PRVT.ZS | Household Final Consumption Expenditure (% of GDP) |
| NV.IND.MANF.ZS | Manufacturing, value added (% of GDP) |
| NY.GDP.MINR.RT.ZS | Mineral Rent (% of GDP) |
| FM.AST.PRVT.GD.ZS | Private Sector Credit (% of GDP) |
| NY.GDP.NGAS.RT.ZS | Natural Gas Rent (% of GDP) |
| GC.NLD.TOTL.GD.ZS | Net Lending (+) / Net Borrowing (-) (% of GDP) |
| NY.GDP.PETR.RT.ZS | Oil Rent (% of GDP) |
| BX.TRF.PWKR.DT.GD.ZS | Personal Remittances, Paid (% of GDP) |
| NV.SRV.TOTL.ZS | Services, value added (% of GDP) |
| NY.GDP.TOTL.RT.ZS | Total Natural Resources Rent (% of GDP) |
| NE.TRD.GNFS.ZS | Trade in Goods and Services (% of GDP) |
| BG.GSR.NFSV.GD.ZS | Net Service Exports (% of GDP) |
| MS.MIL.XPND.ZS | Military Expenditure (% of GNI) |
| MS.MIL.XPND.CD | Military Expenditure (current US$) |
| NV.IND.MANF.KD.ZG | Manufacturing Output Growth (% annual) |
| SP.DYN.LE00.FE.IN | Life Expectancy at Birth, female (years) |
| SP.DYN.LE00.MA.IN | Life Expectancy at Birth, male (years) |
| SH.MMR.RISK.ZS | Maternal Mortality Ratio (modeled estimate, per 100,000 live births) |

| Abbreviation | Full Name |
| --- | --- |
| EG.CFT.ACCS.UR.ZS | Access to Clean Fuels and Technologies for cooking (% of urban population) |
| EG.ELC.ACCS.UR.ZS | Access to electricity (% of urban population) |
| SH.STA.ODFC.UR.ZS | Open Defecation (% of urban population) |
| SH.H2O.BASW.UR.ZS | Basic Water Services (% of urban population) |
| SH.STA.BASS.UR.ZS | Basic Sanitation Services (% of urban population) |
| SH.H2O.SMDW.UR.ZS | Safely Managed Drinking Water Services (% of urban population) |
| SH.STA.SMSS.UR.ZS | Safely Managed Sanitation Services (% of urban population) |
| EN.URB.LCTY.UR.ZS | Urban Population Living in Slums (% of urban population) |
| EN.URB.MCTY.TL.ZS | Urban Population Living in Megacities (% of total) |
| SP.URB.TOTL | Total Urban Population |
| SP.URB.TOTL.IN.ZS | Urban Population (% of total population) |
| SP.URB.GROW | Urban Population Growth (annual %) |
| NE.EXP.GNFS.KD.ZG | Export of Goods and Services (real growth %) |
| NE.CON.TOTL.KD.ZG | Consumption, total (real growth %) |
| NY.GDP.PCAP.CD | GDP per capita (current US$) |
| SI.POV.GAPS | Poverty Gap at National Poverty Lines (%) |
| SI.POV.LMIC.GP | Poverty Gap at $3.20/day (2011 PPP) (%) |
| SI.POV.UMIC.GP | Poverty Gap at $5.50/day (2011 PPP) (%) |
| SI.POV.DDAY | Poverty Headcount Ratio at $1.90/day (2011 PPP) (%) |
| SI.POV.LMIC | Poverty Headcount Ratio at $3.20/day (2011 PPP) (%) |

| Abbreviation | Full Name |
| --- | --- |
| SI.POV.UMIC | Poverty Headcount Ratio at $5.50/day (2011 PPP) (%) |